# Evolution from quartz (**qtz**) to diamond (**dia**) carbon allotropes: Crystal engineering and DFT investigations.

Samir F. Matar

Lebanese German University (LGU), Sahel Alma, Jounieh, Lebanon

https://orcid.org/0000-0001-5419-358X

**Abstract**

*Based on crystal engineering and density functional theory DFT calculations a transformational pathway is proposed from **qtz** (quartz-based) topology characterized by distorted tetrahedra to **dia** (diamond-like) regular tetrahedra topology. The protocol consists of carbon insertions into orthorhombic (space group P222, No. 16) within $C_5$, $C_6$, and $C_7$, leading to ultimate $C_8$ identified as diamond-like. The induced structural and physical changes are assessed with elastic properties pointing to ultra hardness, larger for **qtz**-$C_6$ than **dia**-$C_8$, whilst intermediate $C_7$ is compressible due to its diamond-defective structure. The dynamic stability was shown from the phonons, and thermodynamic quantities as the specific heat $C_V$ was addressed in comparison with diamond experimental data. The electronic band structures reveal semi-conducting $C_6$, metallic $C_7$ characterized by diamond-defect structure, and insulating $C_8$.*

**Introduction**

Searching for original carbon allotropes is an active research field among solid state scientists aiming at achieving mechanical and thermal properties close to diamond. Significant support is provided by modern materials search programs based on evolutionary crystallography used to predict original allotropes [1], with structures subsequently stored in databases [2]. Furthermore, a classification of allotropes into different topologies is helped using TopCryst system [3]. For instance, diamond and related structures with similar corner-sharing (undistorted) tetrahedra are labeled with "**dia**" topology. But structures characterized by distorted tetrahedra, with angles smaller and larger than ∠109.47° ideal value, possess other topologies as detailed herein. The role of the solid-state scientist in such endeavor remains essential, especially in rationalized crystal chemistry criteria and engineering that are validated thanks to quantum mechanical calculations with methods built around the well-established Density Functional Theory (DFT) [4,5]. Such computations allow for accurate assessment of the ground state energy structure and the energy-related properties.

One challenge is raised by an essential property of diamond: its high density of $\rho = 3.53$ g/cm$^3$. Most other carbon allotropes (non-**dia**) have lower densities due to the different packing of *C4* tetrahedra (or for being two-dimensional structures as graphite). But an exceptional feature was observed for a carbon allotrope with quartz-based (**qtz**) topology, hexagonal $C_3$ announced by Luo et al. [6] with the characteristics of being denser than diamond: $\rho = 3.67$ g/cm$^3$. More recently we have proposed hexagonal **qtz** $C_6$ with similar mechanical properties [7]. Another feature accompanying the high density is the distorted *C4* tetrahedron in **qtz** topology with angles below and above the perfect ∠C-C-C sp$^3$ angle of 109.47° (cf. Table 1). The ultimate property is a challenging feature of a larger calculated hardness for **qtz** carbon versus diamond.

In this context the purpose of this communication is to propose a schematic from **qtz** to **dia** topologies following different carbon stoichiometries through progressive carbon atoms insertions, all inscribed in the same orthorhombic space group to enable establishing trends of energy, volume, density, angles, etc. Support of the devised crystal structures is provided by the calculations of the ground state energies and the energy-derived quantities as the elastic properties, the dynamic stability, and the thermodynamic quantities as the specific heat in comparison with experimental data of diamond, as well as the electronic band structures.

### 1- Computational methodology

The identification of the ground state structures corresponding to the energy minima and prediction of their mechanical and dynamical properties were carried out by DFT-based calculations using the Vienna Ab initio Simulation Package (VASP) code [8, 9] and the projector augmented wave (PAW) method [9, 10] for the atomic potentials. Exchange correlation (XC) effects were considered using the generalized gradient functional approximation (GGA) [11]. Relaxation of the atoms onto the ground state structures was performed with the conjugate gradient algorithm according to Press *et al*. [12]. A tetrahedron method was used for geometry optimization and energy calculations [13]. Brillouin-zone (BZ) integrals were approximated by a special **k**-point sampling according to Monkhorst and Pack [15]. Structural parameters were optimized until atomic forces were below 0.02 eV/Å and all stress components were < 0.003 eV/Å$^3$. The calculations were converged at an energy cutoff of 400 eV for the plane-wave basis set in terms of the **k**-point integration in the reciprocal space from $k_x(6) \times k_y(6) \times k_z(6)$ up to $k_x(12) \times k_y(12) \times k_z(12)$ for the final convergence and relaxation to zero strains.

The mechanical stability and hardness were obtained from the calculations of the elastic constants. The treatment of the results was done thanks to ELATE [16] online tool devoted to the analysis of the elastic tensors providing the bulk B and shear G modules along different averaging methods; Voigt's method was used herein [17]. For the calculation of the Vickers hardness a semi-empirical model based on elastic properties was used.

The dynamic stabilities were confirmed from the phonon positive magnitudes. The corresponding phonon band structures were obtained from a high resolution of the Brillouin Zone according to Togo *et al*. [18]. The electronic band structures were obtained using the all-electron DFT-based ASW method [19] and the GGA XC functional [11]. The VESTA (Visualization for Electronic and Structural Analysis) program [20] was used to visualize the crystal structures and charge densities.

### 2- Crystal chemistry

The simplest representation of a tetrahedron in distorted form is depicted in Fig. 1a with ball-and-stick and tetrahedral representations featuring corner sharing irregular tetrahedra (*vide infra*). Central carbon is shown with a white ball within a tetrahedron formed by four carbon atoms (brown) belonging to different sites. This representation results from unconstrained geometry relaxation onto the energy ground state along the methodology described in Section 1. The ground state parameters are given in Table 1 with a weakly cohesive structure characterized in orthorhombic system in space group *P*222 (No. 16) with two Wyckoff positions, namely carbon at body center and C'(4*u*) *x,y,z*. The four-fold (4*u*) atomic position with changing *x,y,z* will be shown to determine the progress of increasing symmetry up to diamond's. A topology analysis of $C_5$ provided a rarely occurring topology: 3,4**L**166, different from the sought **qtz**. The tetrahedron is distorted with angles departing significantly from the ideal value of ∠109.47° and dispersed x, y, z values of (4*u*) position. Such a structure was found little cohesive with an energy of -0.6 eV/atom. Lastly, the density of such an isolated tetrahedron is expectedly low (versus the other carbon allotropes shown in Table 1).

Adding one more carbon atom as shown in Table 1 2$^{nd}$ column leads to orthorhombic $C_6$ shown in Fig. 1b with a structure characterized by corner-sharing, similarly distorted tetrahedra as shown by the angles (~90° <109.47°< 133°). **qtz** topology was identified with TopCrsyt analysis [3]. Orthorhombic $C_6$ was found cohesive with more than twice the magnitude of $C_5$ and a density remarkably large: 3.67 g/cm$^3$ versus ρ(diamond) =3.53 g/cm$^3$ (Table 1).

Going to higher stoichiometry, an additional carbon leading to $C_7$ was inserted at single-multiplicity site in *P*222 space group, resulting into the following carbon positions: (0,0,0), (0,½,½), and (½,0,½). Now the atomic positions at (4*u*) x,y,z become close to ¼, for a fully relaxed structure. Also, the orthorhombic three lattice constants *a, b,* and *c* are now close. Fig. 1c provides a new representation with almost regular tetrahedra characterized by ∠112° and ∠103° slightly higher and lower than 109.47°. The topology analysis provides **bor** topology, adopted by CdIn$_2$Se$_4$ [21] which also contains 7 atoms. The cohesive energy amounts to -1.70 eV/atom, better than with **qtz** $C_6$. The density is lower than in $C_6$ due to the defect structure that can tentatively be described as $C_8$ (diamond stoichiometry) with one carbon vacancy. Lastly, it needs to be mentioned that the reverse operation of removing one carbon from **bor**-$C_7$ followed by geometry relaxations leads back to $C_6$ with **qtz** topology and highly distorted tetrahedra.

The subsequent last step consisted of inserting a fourth carbon atom completing the stoichiometry to $C_8$. The fully relaxed structure can still be described in orthorhombic system *P*222. However, further crystal analysis led to assume a higher cubic symmetry with a single cubic *a* lattice constant of 3.562 Å in face center space group *F-d*3 No. 203 with a unique Wyckoff position of 8*a* (0,0,0). $C_8$ was found characterized with **dia** topology and a cohesive energy of -2.47 eV like diamond. The density is now like diamond, i.e., ρ=3.53 g/cm$^3$, yet lower than ρ(**qtz** $C_6$) =3.67 g/cm$^3$.

Summarizing, the provided schematics depicts a passage from **qtz** to **dia** topologies in systems with corner sharing tetrahedra through progressive carbon insertions causing transformations of distorted tetrahedra in $C_5$, $C_6$ and slightly distorted in $C_7$, all belonging to the same orthorhombic space group into regular undistorted $sp^3$-like ones in $C_8$ stoichiometry with a system described with **dia** topology.

### 3- Physical results and discussions

*3.1 Mechanical properties from the elastic constants*

An analysis of the mechanical properties was carried out with the calculation of the elastic constants $C_{ij}$ (i and j corresponding to directions) by operating finite distortions of the lattice. The system is then fully described by the bulk ($B$) and the shear ($G$) modules obtained by an averaging of $C_{ij}$. The calculated sets of $C_{ij}$ are given in Table 2 for the three carbon stoichiometries. All $C_{ij}$ values are positive letting expect mechanically stable systems. **qtz** $C_6$ and **dia** $C_8$ have the largest $C_{ij}$ values, while intermediate **bor**-$C_7$ exhibits a severe drop of most $C_{ij}$ values versus pristine $C_6$.

The bulk $B_V$ and shear $G_V$ moduli were then obtained from ELATE program [15] for the analysis of the elastic tensors. The last columns of Table 2 provide the obtained $B_V$ and $G_V$ showing magnitudes following the trends observed for $C_{ij}$ with half smaller bulk modulus and lower shear modulus for $C_7$.

Subsequently the hardness according to Vickers ($H_V$) was predicted using the model of Chen et. al. [22] as

$$H_V = 0.92\ (G/B)^{1.137}\ G^{0.708}$$

with $H_V$ possessing the same unit as $G_V$, i.e., GPa.

In this equation the $G/B$ ratio is called the Pugh allowing distinguishing ductile behavior ($G_V/B_V < 1$) from brittle behavior ($G_V/B_V > 1$) [23]. As a result, with $G_V/B_V(C_7) = 0.76$ smaller than 1 is found compressible whereas $G_V/B_V(\mathbf{qtz}\text{-}C_6) = 1.24$ and $G_V/B_V(\mathbf{dia}\text{-}C_8) = 1.19$ are on the brittle side. The last column of Table 2 reflects these observations with the highest magnitude $H_V(\mathbf{qtz}\text{-}C_6) = 102$ GPa pointing to super hard behavior, larger than $H_V(\mathbf{dia}) = 95$ GPa, and lastly the smallest magnitude obtained for intermediate $C_7$ with $H_V = 34$ GPa has a rather soft behavior caused by the defect, rather open structure.

At this point, it needs to be mentioned that considering $C_7$ as template setup, we devised an original carbonitride of well-known stoichiometry $C_3N_4$ (cf. [24] and therein cited works) with the occupation of carbon C' (4u) by N. After geometry optimization to the ground state, a stable ultra-hard **bor**-$C_3N_4$ was identified with systematically higher $C_{ij}$ and resulting ultra hardness $H_V$ =62 GPa (cf. Table 2) almost twice larger than **bor**-$C_7$. These results are likely due to the compensating defect $C_7$ by a binary chemical system with covalent C-N iono-covalent bonding, and the electronic structure property of being isoelectronic of diamond. Indeed, $C_3N_4$ has a valence electron count of 32 (3×4 + 4×5) electrons, identical to $C_8$ (8×4). Of course, this applies for devising silicon nitride $Si_3N_4$, also isoelectronic of $C_3N_4$.

### *3.2 Dynamic and thermodynamic properties from the phonons*

An important criterion of phase dynamic stability is obtained from the phonon's properties. Phonons are quanta of vibrations with their energy quantized with the Planck constant 'h' used in its reduced form ℏ (ℏ = h/2π). The phonons energy is given by E = ℏω where ω is the frequency.

The phonons band structures were obtained following the protocol presented in Section 2. Figure 2 presents the corresponding band structures (red lines) with the horizontal direction showing the main directions of the orthorhombic Brillouin Zone (BZ) while the vertical direction presents the frequencies ω given in units of terahertz (THz). There are 3N-3 optical modes found at higher energy than three acoustic modes that start from zero energy (ω = 0) at the Γ point, BZ center, up to a few Terahertz. They correspond to the lattice rigid translation modes of the crystal (two transverse and one longitudinal). The remaining bands correspond to the optic modes. The absence of any negative frequencies is indicative of dynamically stable systems. Nevertheless, a slightly negative acoustic phonon around BZ center (Γ, Fig. 2a) was assigned to longitudinal vibration along c. However, that doesn't indicate unstable dynamic behavior. Interestingly, the highest optic phonon magnitudes reach 40 THz in both $C_6$ and $C_8$, close to the experimentally observed magnitude observed for diamond by Raman spectroscopy [25], whereas lower magnitudes are observed for intermediate $C_7$. These results support the mechanical properties and the mechanical closeness of $C_6$ to $C_8$.

The thermodynamic properties of the new allotropes were calculated from the phonon frequencies using the statistical thermodynamic approach [26] on a high-precision sampling mesh in the orthorhombic Brillouin zone. The temperature dependencies of the heat capacity at constant volume ($C_V$) are shown in Figure 3 in comparison with experimental $C_V$ data formerly obtained for diamond [27]. The experimental points coincide well with $C_8$ curve of $C_V$, followed by $C_6$, and lastly $C_7$. Then thermally, $C_6$ is closer to $C_8$ than $C_7$.

*3-3 Electronic band structures*

Using the crystal parameters in Table 1, the electronic band structures were obtained using the all-electrons DFT-based augmented spherical method (ASW) [19] using GGA XC functional. The bands are shown in Figure 4. Along the horizontal axis the bands develop along the main directions of the orthorhombic BZ. For $C_6$ and $C_8$, the zero energy along the vertical axis is considered with respect to $E_V$, at the top of the filled valence band (VB) separated from the empty conduction band (CB) by an indirect energy gap along Γ-X in both allotropes with ~2 eV in $C_6$ and twice amore in diamond-like $C_8$. Oppositely, diamond-defect $C_7$ is metallic with bands crossing the Fermi level $E_F$.

## 4- Conclusion

The main purpose of this paper was to highlight a relationship between two hard carbon allotropes adopting **qtz** and **dia** topologies through a progressive increase of stoichiometries from elementary tetrahedron to **qtz** $C_6$, then to intermediate stoichiometry **bor**-$C_7$, and lastly to **dia**-$C_8$. Such a study was made possible through devising these original carbon allotropes in a low symmetry orthorhombic space group, namely *P*222 No.16 allowing for several single-occupations Wyckoff positions beside a single four-fold one. A major observation was found in the changes brought by the addition of carbon into distorted tetrahedron structure $C_5$ going progressively from highly distorted in **qtz** $C_6$ to regular tetrahedra in **dia**-$C_8$, with intermediate distortion in **bor**-$C_7$. The crystal engineering manipulations were backed with geometry unconstrained crystal optimizations down to ground state energy structures followed by deriving quantities as the elastic constants providing the bulk and shear modules as well as the phonons band structures and depending thermodynamic properties. The Vickers hardness ($H_V$) magnitudes point to ultra hardness of **qtz**-$C_6$ with $H_V$=102 GPa, higher than **dia**-$C_8$ ($H_V$=95 GPa), whilst **bor**-$C_7$ was found highly compressible due to defect structure. The studied systems were found with dynamic stabilities from deriving the phonons expressed in band structures. The calculated respective specific heats $C_V$ were compared with experimental data of diamond. The electronic band structures point to semi-conducting $C_6$, insulating $C_8$ and metallic $C_7$.

*Conflict of Interest*: The author declares no conflict of interest.

Table 1 Crystal structure parameters of carbon allotropes with different **topologies**.

| Topology<br>*P*222 No. 16 | 3,4**L**166<br>$C_5$ (single distorted tetrahedron) | **qtz**<br>$C_6$ | **bor**<br>$C_7$ | **dia**<br>$C_8$ | **dia**<br>$C_8$ *Fd*-3 No.203 |
|---|---|---|---|---|---|
| *a*, Å | 2.829 | 2.793 | 3.550 | 3.5621 | 3.56 |
| *b*, Å | 5.183 | 4.499 | 3.572 | 3.5623 | |
| *c*, Å | 2.517 | 2.597 | 3.543 | 3.5621 | |
| $V_{cell}$, Å³ | 36.91 | 32.63 | 44.92 | 45.20 | 45.12 |
| $<V_{atom}>$ Å³ | 7.38 | 5.44 | 6.42 | 5.65 | 5.64 |
| *Density* g/cm³ | 2.74 | 3.67 | 3.12 | 3.53 | 3.53 |
| *Shortest d-d-* Å | 1.62 | 1.59 | 1.50 | 1.54 | 1.54 |
| Angles (deg.) | ∠C-C'-C 96.5°/134° | ∠C'-C1-C' 90.12°/133° | ∠C'-C1-C' 112.68°/102.92° | ∠109.47° | ∠109.47° |
| Atomic position | C (1c) ½, ½, ½<br>C' (4u) 0.868,0.289,0.247 | C1 (1a) 0,0,0<br>C2 (1g) 0,½,½<br>C' (4u) 0.334,0.250,0.244 | C1 (1a) 0,0,0<br>C2 (1f) ½,0,½<br>C3 (1g) 0,½,½<br>C' (4u) 0.234,0.266,0.234 | C1 (1a) 0,0,0<br>C2 (1f) ½,0,½<br>C3 (1g) 0,½,½<br>C4 (1e) ½ ½,0<br>C' (4u) ¼ ¼ ¼ | C(8a) 0,0,0 |
| E(coh.)/at. eV | -0.60 (unstable) | -1.32 | -1.70 | -2.47 | -2.47 |

N.B. E(C) = -6.6 eV. $E_{diamond}$ (coh.)/at. = -2.49 eV.

Table 2. Carbon allotropes elastic constants, bulk $B_V$, and shear $G_V$ moduli and hardness $H_V$ in GPa units. *The results obtained for $C_7$-derived $C_3N_4$ with same topology are shown for sake of argumentation*

| *P*222 | $C_{11}/C_{22}$ | $C_{12}$ | $C_{13}/C_{23}$ | $C_{33}$ | $C_{44}$ | $C_{55}$ | $C_{66}$ | $B_V$ | $G_V$ | $H_V$ |
|---|---|---|---|---|---|---|---|---|---|---|
| **qtz** $C_6$ | 1164/1186 | 63 | 63/88 | 1188 | 540 | 544 | 540 | 441 | 547 | 102 |
| **bor** $C_7$ | 613/614 | 182 | 182/181 | 614 | 271 | 271 | 270 | 326 | 249 | 34 |
| ***bor** $C_3N_4$- from $C_7$* | *845/854* | *223* | *223/209* | *841* | *456* | *466* | *470* | *423* | *406* | *62* |
| **dia** $C_8$ | 1068/1068 | 133 | 133/133 | 1068 | 567 | 567 | 572 | 445 | 528 | 95 |

FIGURES

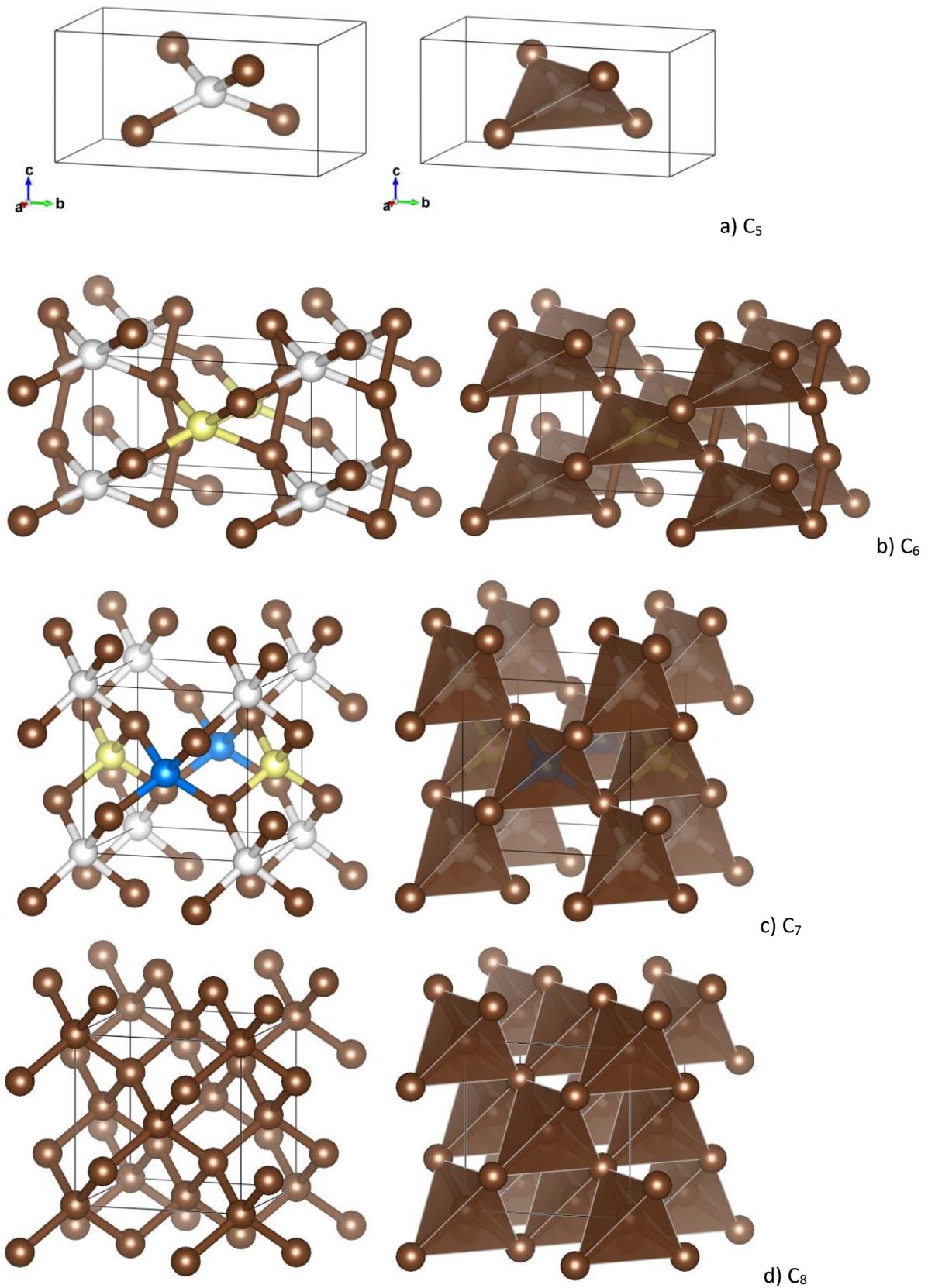

a) C₅

b) C₆

c) C₇

d) C₈

Figure 1. Crystal structure representations with ball and stick (left) and polyhedral (right). The colored spheres symbolize the addition of carbon atoms described in Table 1. In highest symmetry C₈, all carbon atoms are identical.

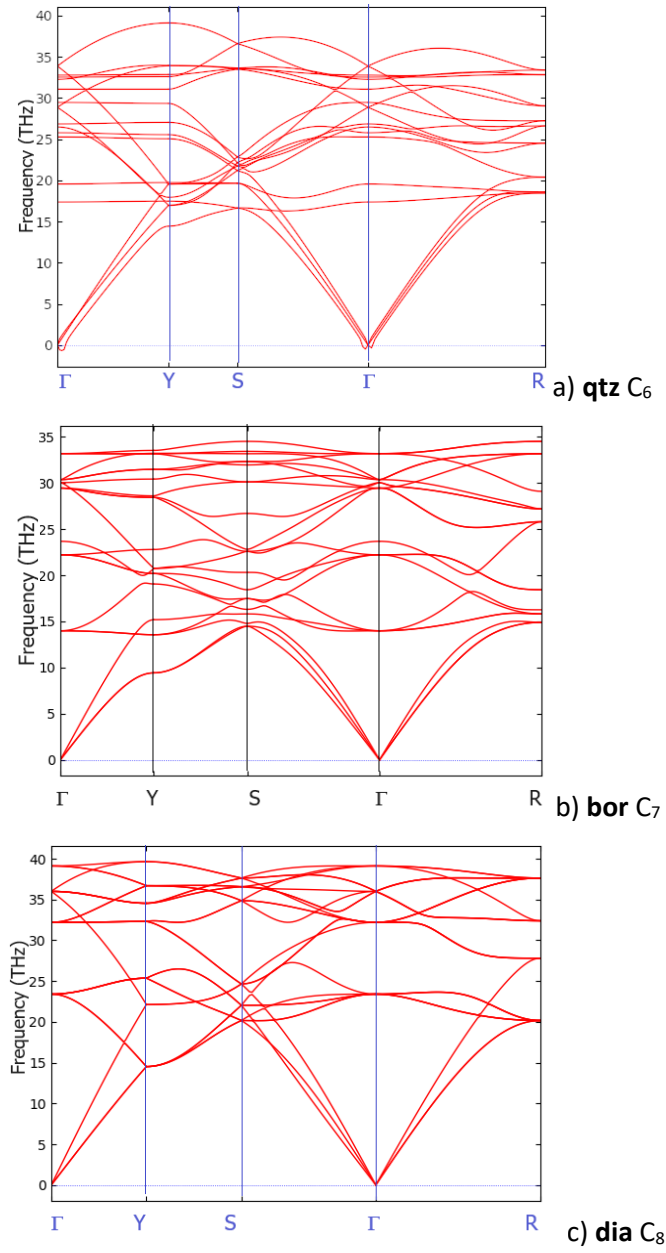

Figure 2. Phonons band structures along orthorhombic Brillouin zone.

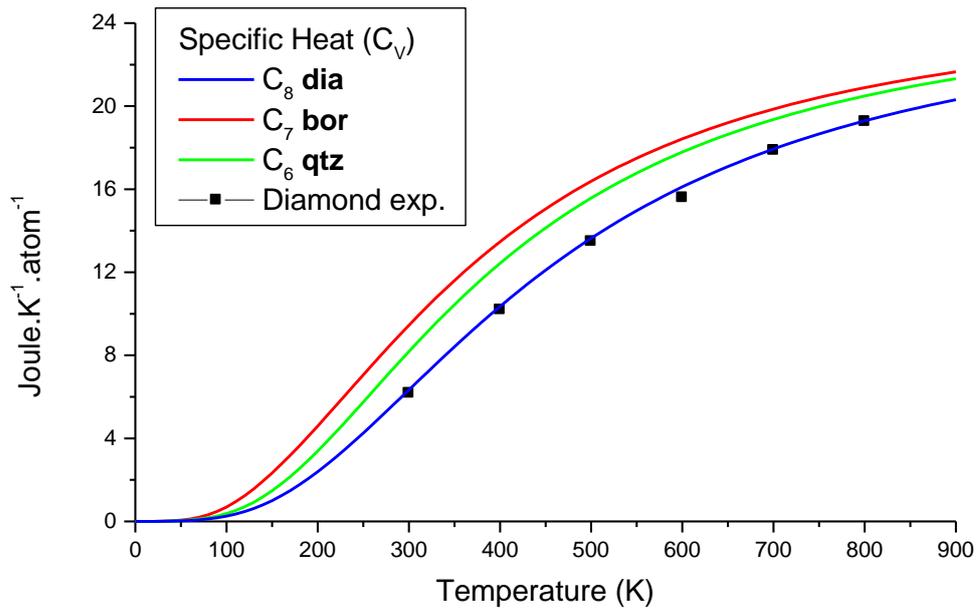

Figure 3. Temperature changes of the specific heat $C_V$.

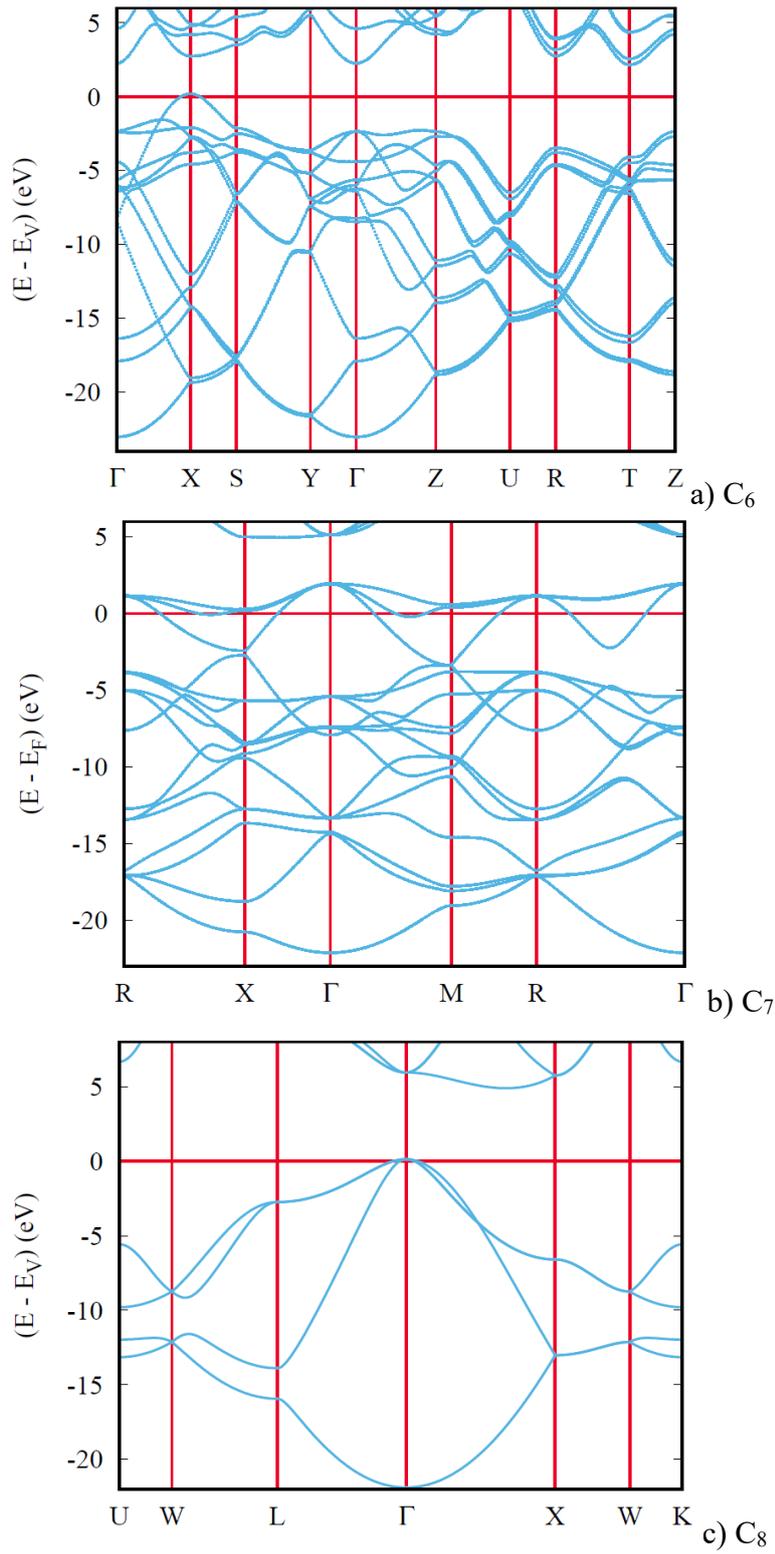

Figure 4. Electronic band structures.